# Atom Localization in two and three dimensions via level populations

Nilesh Chaudhari and Amarendra K. Sarma*
Department of Physics, Indian Institute of Technology Guwahati, Guwahati-781039, Assam, India.
*aksarma@iitg.ernet.in

ABSTRACT: We propose schemes for two-dimensional (2D) and three-dimensional (3D) atom localization in a five-level M-type system using standing-wave laser fields. In the upper two levels of the system we see a 'coupled' localization for both 2D and 3D case. Here, the state in which majority of population will be found depends on the sign of the detunings between the upper levels and the intermediate level. The experimental implementation of the scheme using the $D_2$ line of $^{87}$Rb is also proposed. The scheme may be manipulated to achieve subwavelength localization of atoms in one dimension to a spatial width, smaller by a factor of 1000 than the incident wavelength.

## 1. Introduction

In recent years atom localization has received considerable attention owing to its potential applications in the areas of quantum information science [1], laser cooling and trapping of neutral atoms [2], Bose-Einstein condensation [3], atom nanolithography [4, 5] and microscopy [6]. Initial proposals for localizing atoms were confined to one dimension only [7-17]. In this regard, many methods based on quantum interference [7], electromagnetically induced transparency (EIT) [1,8], coherent population trapping (CPT)[9,10] and stimulated Raman adiabatic passage (STIRAP) [11] etc. are suggested. Recently, schemes have been proposed to localize atoms even in two and three dimensions. Ivanov and Rozhdestvensky have proposed a scheme for two dimensional (2D) atom localization by laser fields in a four-level tripod system [18]. Very recently, Ivanov et al. have proposed a scheme for three dimensional atom localization in the same system [19]. It should be noted that other schemes for 2D localization have also been proposed [20-23]. On the other hand, Qi et al. [24] have proposed a scheme for 3D atom localization based on EIT. Clearly these new developments are going to open up numerous applications in many areas of science. For example, three-dimensional atom localization is speculated to be useful in high-precision position dependent chemistry [19]. Few experimental developments have already been made towards localizing atoms [4,25,26]. Johnson et al. reported localization of metastable atom beams and opened the door for nanolithography at the Heisenberg limit [4]. One notable progress in experimental realization of atom localization in one dimension is made by Yavuz's group [26]. They have reported, using EIT, experimental realization of subwavelength localization of atoms to a spatial width smaller by a factor of eight to the incident wavelength. It is expected that many more experiments will be carried out soon and we need practical schemes to achieve the goal of atom localization in all the dimensions. In this work, we propose schemes to localize atoms both in two and three dimensions, via level populations, by using laser fields in a five level M-type atomic system. Our schemes are not only useful to localize atoms in two and three dimensions, but also enable one to achieve subwavelength localization of atoms in one dimension to a spatial width, smaller by a factor of 1000 than the incident wavelength. This article is organized as follows. In Section 2 we introduce our model and give the basic equations. Section 3 contains our simulated results and discussions followed by conclusions in Section 4.

## 2. Theoretical Model

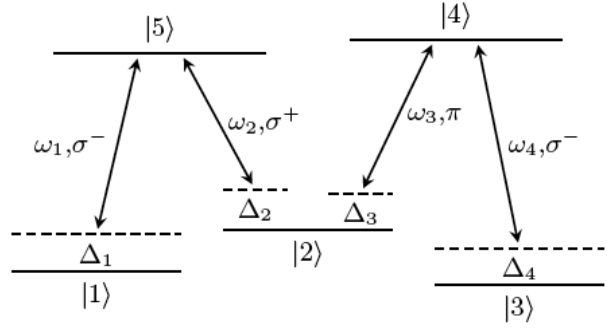

Fig. 1. Schematics of the M-type atomic system

We consider an M-type atomic system as depicted in Fig. 1. The four laser fields $\vec{E_1}, \vec{E_2}, \vec{E_3}$ and $\vec{E_4}$ form a standing wave pattern in a plane and atoms cross the plane in a direction perpendicular to the plane. The electric field pattern is given by [27]:

$$\vec{E}(\vec{r},t) = \vec{E_1^+}(x)e^{-i\omega_1 t} + \vec{E_2^+}(y)e^{-i\omega_2 t} + \vec{E_3^+}(y)e^{-i\omega_3 t} + \vec{E_4^+}(x)e^{-i\omega_4 t} + cc. \tag{1}$$

Here $\vec{E_1}$ and $\vec{E_4}$ are $\sigma^-$ polarized while $\vec{E_2}$ is $\sigma^+$ polarized and $\vec{E_3}$ is $\pi$ polarized wave. $\omega_j$ is the frequency of the field $\vec{E_j}$. As shown in Fig. 1, the upper levels are denoted by $|4\rangle$ and $|5\rangle$, and the three ground levels by $|1\rangle, |2\rangle$ and $|3\rangle$. The transitions $|1\rangle \leftrightarrow |5\rangle, |2\rangle \leftrightarrow |5\rangle, |2\rangle \leftrightarrow |4\rangle, |3\rangle \leftrightarrow |4\rangle$ are driven by nearly resonant electric fields $\vec{E_1}(x) = \vec{E_1}\sin(x), \vec{E_2}(y) = \vec{E_2}\sin(y), \vec{E_3}(y) = \vec{E_3}\sin(y), \vec{E_4}(x) = \vec{E_4}\sin(x)$ respectively. The detunings for these transitions are: $\Delta_1 = \omega_1 - \omega_{51}, \Delta_2 = \omega_2 - \omega_{52}, \Delta_3 = \omega_2 - \omega_{42}$ and $\Delta_4 = \omega_4 - \omega_{43}$ respectively. The coupling of the laser fields with the atom is given by Rabi frequencies [28]:

$$\begin{aligned} g_1(x) &= G_1 \sin(k_1 x) \\ g_2(y) &= G_2 \sin(k_2 y) \\ g_3(y) &= G_3 \sin(k_3 y) \\ g_4(x) &= G_4 \sin(k_4 x) \end{aligned} \tag{2}$$

where $G_j = |\vec{d_j}.\vec{E_j}/\hbar|$ is the coefficient of Rabi frequency and $\vec{d_j}$ is the dipole moment corresponding to the j-th transition. It is assumed that the center of mass of the atom is at rest, the interaction only affects the internal states and hence the Raman–Nath approximation is valid [29]. So in the interaction picture and with the rotating wave approximation (RWA) [30], the Liouville equation:

$$i\hbar\dot{\rho} = [H,\rho] - i\gamma\rho \tag{3}$$

where $\rho$ is the density matrix operator and $H$ is the Hamiltonian of the system, becomes:

$$i\dot{\rho}_{11} = g_1(\rho_{15} - \rho_{51}) + i\gamma_1\rho_{55}, \tag{4a}$$

$$i\dot{\rho}_{22} = g_3(\rho_{24} - \rho_{42}) + g_2(\rho_{25} - \rho_{52}) + i\gamma_{52}\rho_{55} + i\gamma_{42}\rho_{44}, \tag{4b}$$

$$i\dot{\rho}_{33} = g_4(\rho_{34} - \rho_{43}) + i\gamma_3\rho_{44}, \tag{4c}$$

$$i\dot{\rho}_{44} = g_3(\rho_{42} - \rho_{24}) + g_4(\rho_{43} - \rho_{34}) - i\gamma_4\rho_{44}, \tag{4d}$$

$$i\dot{\rho}_{55} = g_1(\rho_{51} - \rho_{15}) + g_2(\rho_{52} - \rho_{25}) - i\gamma_5\rho_{55}, \tag{4e}$$

$$i\dot{\rho}_{12} = \rho_{12}(\Delta_{12} - i\Gamma_{23}) + g_3\rho_{14} + g_2\rho_{15} - g_1\rho_{52}, \tag{4f}$$

$$i\dot{\rho}_{13} = g_4\rho_{14} - g_1\rho_{53} + (\Delta_{1234} - i\Gamma_{13})\rho_{13}, \tag{4g}$$

$$i\dot{\rho}_{14} = \rho_{14}(\Delta_{123} - i\Gamma_{14}) + g_3\rho_{12} + g_4\rho_{13} - g_1\rho_{54}, \tag{4h}$$

$$i\dot{\rho}_{15} = g_1(\rho_{11} - \rho_{55}) + g_2\rho_{12} + \rho_{15}(\Delta_1 - i\Gamma_{15}), \tag{4i}$$

$$i\dot{\rho}_{23} = \rho_{23}(\Delta_{43} - i\Gamma_{23}) - g_3\rho_{43} - g_2\rho_{53} + g_4\rho_{24}, \tag{4j}$$

$$i\dot{\rho}_{24} = g_3(\rho_{22} - \rho_{44}) + g_4\rho_{23} - g_2\rho_{54} + \rho_{24}(\Delta_3 - i\Gamma_{24}), \tag{4k}$$

$$i\dot{\rho}_{25} = g_2(\rho_{22} - \rho_{55}) + g_1\rho_{21} - g_3\rho_{45} + \rho_{25}(\Delta_2 - i\Gamma_{25}), \tag{4l}$$

$$i\dot{\rho}_{34} = g_4(\rho_{33} - \rho_{44}) + g_3\rho_{32} + \rho_{34}(\Delta_4 - i\Gamma_{34}), \tag{4m}$$

$$i\dot{\rho}_{35} = \rho_{35}(\Delta_{234} - i\Gamma_{35}) + g_1\rho_{31} + g_2\rho_{32} - g_4\rho_{45}, \tag{4n}$$

$$i\dot{\rho}_{45} = \rho_{45}(\Delta_{23} - i\Gamma_{45}) + g_1\rho_{41} + g_2\rho_{42} - g_3\rho_{25} - g_4\rho_{35}, \tag{4o}$$

$\rho_{ij}$ are the matrix elements of the density matrix operator, $\rho$, with $\rho_{ij} = \rho_{ji}^*$, $\sum \rho_{ii} = 1$. For $\Delta$s we have used the following notation: $\Delta_{ijkl} = \Delta_i - \Delta_j + \Delta_k - \Delta_l$, $\Delta_{xyz} = \Delta_x - \Delta_y + \Delta_z$ and $\Delta_{mn} = \Delta_m - \Delta_n$. $\gamma_1$ and $\gamma_3$ are the decay rates which corresponds to the relaxation into the ground states $|1\rangle$ and $|3\rangle$ respectively. $\gamma_{52}$ is the decay rate that corresponds to the relaxation from excited state $|5\rangle$ into the ground state $|2\rangle$, similarly other $\gamma$s are defined. $\Gamma_{34}$ corresponds to the decay between states $|3>$ and $|4>$, similarly other $\Gamma$s are defined. The decay rate between states $|1\rangle$ and $|2\rangle$, $\Gamma_{12}$, can safely be neglected because there is no field driving the transition and is usually much smaller than the decay rates corresponding to driven transitions. Similarly other decay rates can be neglected. In effect, we have:

$$\Gamma_{12} \approx \Gamma_{13} \approx \Gamma_{14} \approx \Gamma_{23} \approx \Gamma_{35} \approx \Gamma_{45} \approx 0. \tag{5}$$

## 3. Results and Discussion

### 3.1 2D Localization

We consider the two probe fields $g_2$ and $g_3$ to be sufficiently weak compared to quantities of interest such as $g_1, g_4, \Delta_2, \Delta_3, \Gamma_{24}$ and $\Gamma_{25}$. In the long time limit, we have $\dot{\rho}_{ij} = 0$. From these considerations we can obtain $\rho_{44}, \rho_{55}$ from Eq. (4), after few algebraic manipulations:

$$\rho_{55} \approx \frac{2g_2^2 \Gamma_{25}}{\gamma_0 \left(\Gamma_{25}^2 + \left(\frac{g_1^2}{\Delta_0} - \frac{g_3^2}{\Delta_{23}} + \Delta_2\right)^2\right)} \tag{6a}$$

$$\rho_{44} \approx \frac{2g_3^2 \Gamma_{24}}{\gamma_0 \left(\Gamma_{24}^2 + \left(\frac{g_4^2}{\Delta_0} + \frac{g_2^2}{\Delta_{23}} + \Delta_3\right)^2\right)} \tag{6b}$$

Eq. (6a) is valid near nodes of $g_2$ and away from nodes of $g_3$, while Eq. (6b) is valid near nodes of $g_3$ and away from nodes of $g_2$. In obtaining the above equations, for simplicity, we have taken: $\gamma_5 - \gamma_1 = \gamma_4 - \gamma_3 = \gamma_0$, $\Delta_{12} = \Delta_{34} = \Delta_0$.

Fig. 2 depicts the populations in states $|4\rangle$ and $|5\rangle$ as a function of $(x, y)$. We have used the following parameters [18]: $\Gamma_{24} = 1.6\gamma_0, \Gamma_{25} = 1.4\gamma_0$, $g_2 = 4\gamma_0\sin(k_2 y)$, $g_3 = 4\gamma_0\sin(k_3 y)$, $g_1 = 6\gamma_0\sin(k_1 x)$, $g_4 = 6\gamma_0\sin(k_4 x)$, $\Delta_0 = -6\gamma$, $\Delta_{23} = -2\gamma$, $\Delta_2 = 15\gamma$ and $\Delta_3 = 17\gamma$.

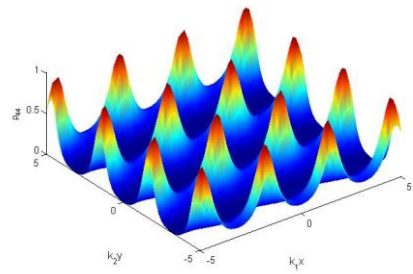

(a)

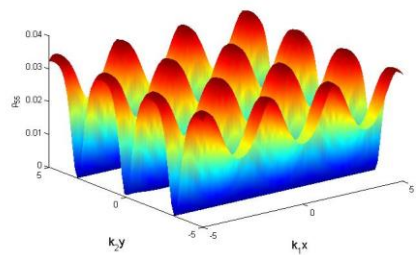

(b)

Fig. 2 Populations in level (a) $|4\rangle$ and (b) $|5\rangle$ as a function of $(x, y)$.

It can be seen that most of the excited population is localized in a single excited state depending on the sign of the detuning $\Delta_{23}$. On the other hand, if $g_2$ is of the form: $g_2 = G_2\cos(k_2 y)$ while other three fields being of the form given in Eq. (2), we obtain a localization pattern where both the excited states are almost equally populated with different localization structures as shown in Fig. (3).

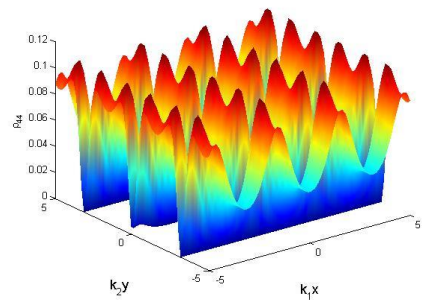

(a)

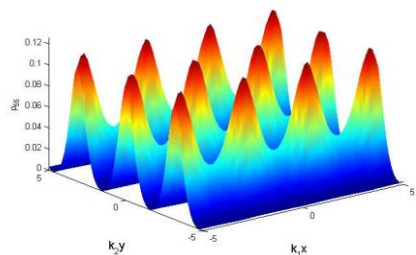

(b)

Fig. 3 Phase shifted populations in level (a) $|4\rangle$ and (b) $|5\rangle$ as a function of $(x, y)$. Here $g_2$ is a cosine wave.

We find that localization of populations crucially depends on the sign of $\Delta_{23}$, i.e. on the sign of the difference between detuning $\Delta_2$ and $\Delta_3$. For a negative value of $\Delta_{23}$, the excited state population is localized primarily in state $|4\rangle$. On the other hand for $\Delta_{23} > 0$, population is tightly and highly localized in state $|5\rangle$.

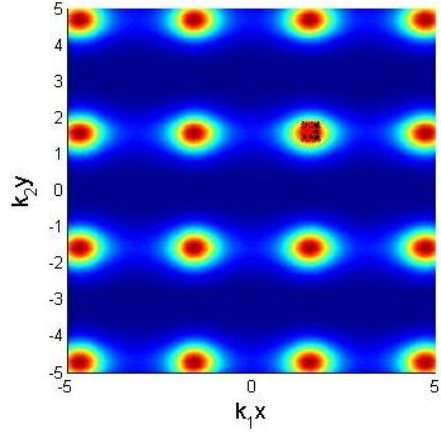

Fig. 4 Contour plot of Fig.(2a): large population can be tightly localized in state $|4\rangle$.

Fig. 4 depicts the corresponding contour plot of Fig. 2(a). From this plot one can estimate the range of localization area of the atomic populations. We estimate that a significant population could be very tightly localized up to a subwavelength region having an area on the order of $A_{min} \approx \left(\frac{\lambda}{300\pi}\right)^2$ in the state $|4\rangle$. It is worthwhile to note that, through judicious manipulation of the proposed scheme, we may achieve subwavelength localization of atoms in one dimension to a spatial width by a factor of 1000 smaller than the wavelength of the laser beam via level population in the state $|4\rangle$.

### 3.2 3D Localization

The above scheme for localization of atoms in two dimensions can easily be modified to localize atoms in three dimensions. The atomic states arrangement is similar to the one with 2D localization except that now two additional fields are required whose $\hat{k}$ vectors are along $\hat{z}$ direction and need to satisfy certain selection rules. The schematic of the proposed scheme is shown in Fig. 5.

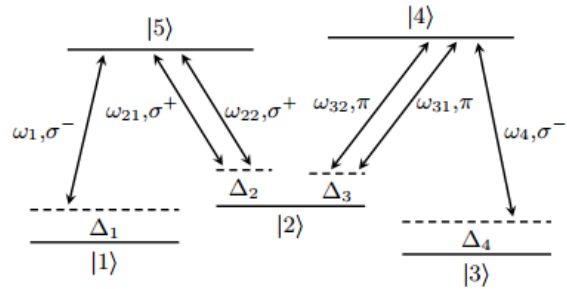

Fig. 5: Configuration of electric fields for an M-type system for 3D localization.

The electric field configuration is of the form:

$$\vec{E}(\vec{r},t) = \overrightarrow{E_1^+}(x)e^{-i\omega_1 t} + \overrightarrow{E_{21}^+}(y)e^{-i\omega_{21} t} + \overrightarrow{E_{22}^+}(z)e^{-i\omega_{22} t} + \overrightarrow{E_{31}^+}(y)e^{-i\omega_{31} t} +$$

$$\overrightarrow{E_{32}^+}(z)e^{-i\omega_{32}t} + \overrightarrow{E_4^+}(x)e^{-i\omega_4t} + cc. \tag{7}$$

The Rabi frequencies associated with the electric field configuration are given by:

$$\begin{aligned}
g_1(x) &= G_1 \sin(k_1 x) \\
g_2(y,z) &= G_{21}\sin(k_{21}y) + iG_{22}\sin(k_{22}z) \\
g_3(y,z) &= G_{31}\sin(k_{31}y) + iG_{32}\sin(k_{32}z) \\
g_4(x) &= G_4 \sin(k_4 x)
\end{aligned} \tag{8}$$

The peak Rabi frequencies $G_{21}, G_{22}, G_{31}, G_{32}$ are defined as: $G_{ij} = |\vec{d}_i \cdot \vec{E}_{ij}/\hbar|$. We solve the Liouville equation, in the long time limit, to obtain:

$$\rho_{44} \approx \frac{2|g_3^2|\Gamma_{24}}{(\gamma_4-\gamma_3)(\Gamma_{24}^2+(\frac{g_4^2}{\Delta_{34}}+\frac{|g_2^2|}{\Delta_{23}}+\Delta_3)^2)}. \tag{9a}$$

$$\rho_{55} \approx \frac{2|g_2^2|\Gamma_{25}}{(\gamma_5-\gamma_1)(\Gamma_{25}^2+(\frac{g_1^2}{\Delta_{12}}-\frac{|g_3^2|}{\Delta_{23}}+\Delta_2)^2)} \tag{9b}$$

In obtaining the solutions we have used the interaction picture and employed the so-called rotating wave approximation. Fig. 6 depicts the 3D plot for atom localization in the state 4 >. Following parameters are used in obtaining Fig. 6: $g_3 = 0.3\gamma$, $g_2^2 = 16\gamma^2(\sin^2(k_{21}y) + \sin^2(k_{22}z))$, $g_4 = 6\gamma \sin(k_4 x)$, $g_1 = 6\gamma \sin k_1 x$, $\Delta_{34} = 9\gamma$, $\Delta_{23} = 4\gamma$, $\Delta_3 = -12\gamma$, $\Gamma_{25} = \Gamma_{24} = 1.5\gamma$ and $\gamma_5 - \gamma_1 = \gamma_4 - \gamma_3 = \gamma$.

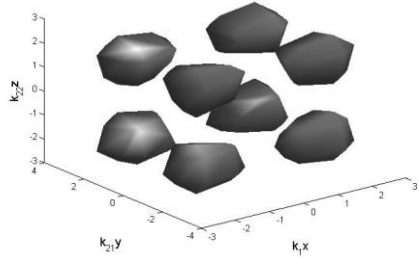

Fig. 6. 3D localization of population |4 >

It can be observed that the 3D structures are periodic in nature. We find that the shape of these structures depend crucially on the atom-field coupling via detuning parameters, $\Delta_{23}$ and $\Delta_{34}$. It is quite evident from Eq. 9(a) also.

The proposed 3D atom localization scheme may be implemented experimentally using the $D_2$ line of [87]Rubidium atom, with various atomic states as described by Eq. (10) below:

$$\begin{aligned}
|1\rangle &= |5^2 S_{1/2}, F = 1, m = +1\rangle, \\
|2\rangle &= |5^2 S_{1/2}, F = 1, m = -1\rangle, \\
|3\rangle &= |5^2 S_{1/2}, F = 1, m = 0\rangle, \\
|4\rangle &= |5^2 P_{3/2}, F' = 1, m = -1\rangle, \\
|5\rangle &= |5^2 P_{3/2}, F' = 0, m = 0\rangle.
\end{aligned} \tag{10}$$

The polarization of the laser fields driving these transitions should be in accordance with the following selection rules:

$\sigma^+$ , if  $m_f - m_i = +1$
$\pi$, if  $m_f - m_i = 0$  (11)
$\sigma^-$ , if  $m_f - m_i = -1$

It is worthwhile to note that by switching off the z- dependent fields, we can achieve atom localization strictly in two dimensions. Further, from 2D we can obtain localization in 1D as well, however localization strictly in 1D is not possible.

## 4. Conclusions

In conclusion, we have shown how to localize atoms in an M-type system in two as well as three dimensions. The most striking feature of the scheme is that one can obtain atom localization in all the dimensions, i.e. 1D, 2D and 3D with judicious manipulations of the applied electric fields. However, localization of atoms only and only in one dimension is not possible. It is also pointed out that for 2D localization, the state in which majority of the population resides depends crucially on the sign of detuning $\Delta_{23}$. We also estimated the range of localization numerically. The experimental implementation of the scheme using the $D_2$ line of [87]Rb is also proposed. The scheme may be manipulated to achieve subwavelength localization of atoms in one dimension to a spatial width, smaller by a factor of 1000 than the incident wavelength.

**ACKNOWLEDGMENTS**
A.K.S. would like to acknowledge the financial support from CSIR (Grant No. 03(1252)/12/EMR-II).